\documentclass[reprint,amsmath,amssymb,aps,prl,superscriptaddress]{revtex4-1}

\usepackage{times,color,amsmath}
\usepackage{subfigure,epsfig,graphicx,epstopdf}
\usepackage[none]{hyphenat}
\usepackage[colorlinks]{hyperref}
\hypersetup{linkcolor = {blue},	citecolor = {blue},	urlcolor = {blue}}

\newcommand{\afft}{\thanks{These authors contributed equally to this work}}
\newcommand{\affa}{\affiliation{Department of Applied Physics, KTH Royal Institute of Technology, Albanova University Centre, Roslagstullsbacken 21, 106 91 Stockholm, Sweden}}
\newcommand{\affb}{\affiliation{Naturwissenschaftlich-Technische Fakultät, Universität Siegen, Walter-Flex-Straße 3, D-57068 Siegen, Germany}}
\newcommand{\affc}{\affiliation{National Research Council of Canada, Ottawa, Ontario, Canada, K1A 0R6}}

\begin{document}

\title{ Scalable generation and detection of on-demand W states in nanophotonic circuits}

\author{Jun Gao} \email{junga@kth.se} \afft   \affa   
\author{Leonardo Santos} \afft   \affb 
\author{Govind Krishna}\email{govindk@kth.se}  \afft \affa
\author{Ze-Sheng Xu} \affb \affa
\author{Adrian Iovan} \affa 	
\author{Stephan Steinhauer} \affa
\author{Otfried Gühne}   \affb 
\author{Philip J. Poole} \affc 	
\author{Dan Dalacu} \affc 	
\author{Val Zwiller} \affa 	
\author{Ali W. Elshaari} \email{elshaari@kth.se} \affa 

\date{\today}

\begin{abstract}

Quantum physics phenomena, entanglement and coherence, are crucial for quantum information protocols, but understanding these in systems with more than two parts is challenging due to increasing complexity. The W state, a multipartite entangled state, is notable for its robustness and benefits in quantum communication. Here, we generate an 8-mode on-demand single photon W states, using nanowire quantum dots and a silicon nitride photonic chip. We demonstrate a reliable, scalable technique for reconstructing W-state in photonic circuits using Fourier and real-space imaging, supported by the Gerchberg-Saxton phase retrieval algorithm. Additionally, we utilize an entanglement witness to distinguish between mixed and entangled states, thereby affirming the entangled nature of our generated state. The study provides a new imaging approach of assessing multipartite entanglement in W-states, paving the way for further progress in image processing and Fourier-space analysis techniques for complex quantum systems.

\end{abstract}

\maketitle

Correlations form the basis for scientific inferences about the world. One of the most notable examples is that of causal inference where correlations between events are explained in terms of models that relate them from direct causation and/or shared common cause\cite{pearl2009causality}. This is a central paradigm in data analysis in science (e.g, cosmology, medical and social sciences), whose results impact from our own understanding of reality to decision making in public policies. In all these cases, probabilities (and consequently, correlations) arise due to ignorance about all the parameters behind the analyzed events. In contrast, entanglement is a particular type of correlation between space-like separated quantum systems for which there is no counterpart in the classical world. This fact is precisely stated by Bell's theorem\cite{bell1964einstein}, which demonstrates the impossibility of reproducing correlations between measurement results performed on entangled quantum systems in terms of local-hidden-variable models, a result whose experimental verification and impact on quantum information science led to the 2022 Nobel Prize in Physics.

The characterization and detection of entanglement, as well as their impact in subsequent generation and experimental manipulation, is therefore of paramount importance. Such questions, however, are equally challenging, both theoretically and experimentally. These difficulties are particularly accentuated when we consider entanglement between more than two particles, here called multipartite entanglement. A fundamental problem lies in the exponential scaling of the dimension of the underlying Hilbert space, thus rendering an exhaustive classification difficult. In order to gain insight into multipartite entanglement phenomena, different concepts based on symmetries, graphical representations, matrix-product approximations, etc., have been used to select quantum states with particularly relevant properties within some context (see, e.g., Refs. \cite{horodecki2009,bengtsson2017,guhne2009}). In this work we are interested in the so-called W states of $N$ qubit systems. These states are characterized by a coherent superposition of all the qubits involved, with equal probability amplitudes. They gained prominence in the scientific literature in the context of multipartite entanglement classification\cite{dur2000}. As it turns out, such states are intrinsically robust against particle loss and have been shown to be central as a resource in quantum information processing and multiparty quantum communication \cite{sohbi2015,barnea2015,murao1999,shi2002,joo2003,guhne2009,agrawal2006perfect}. Furthermore, W states are examples of the so-called Dicke states, which are quantum states that arise naturally in the study of the emission of light by a cloud of atoms via so-called superradiance \cite{dicke1954coherence}.

In the last two decades, the precise control of quantum systems allowed the experimental generation of multipartite entangled states \cite{erhard2020}. Several schemes have been presented for preparing W states in a variety of physical platforms, including cavity quantum electrodynamics \cite{guo2002scheme,zang2016}, quantum spin chains \cite{wang2001entanglement}, nuclear magnetic resonance \cite{vandersypen2005nmr,dogra2015,das2015}, atomic systems \cite{haas2014entangled,hosten2016measurement,mcconnell2015entanglement,frowis2017experimental,pu2018experimental,li2021multipartite} and trapped ions \cite{roos2004,haffner2005scalable}. These schemes are frequently not scalable and/or require complex quantum state witnesses or quantum state tomography for their analysis. Single photons in optical platforms, in contrast, can be generated and manipulated with a high degree of purity, which makes them promising candidates for high-order W state generation. The generation of W states on such platforms, however, is yet challenging since it typically requires complex bulk-optical set-ups\cite{papp2009characterization,choi2010}.

In this work, we propose a scalable method for generating and detecting W states in nanophotonic circuits. We experimentally generate an 8-mode W state on an integrated nanophotonic circuit based on cascaded arrays of Y-branch splitters. The circuit is fabricated on a complementary-metal-oxide semiconductor (CMOS) compatible silicon nitride platform. On-demand single photons generated from a InAsP nanowire quantum dot are fibre-coupled onto the photonic chip with the nanophotonic circuit. The output facet of the chip is imaged, generating real and Fourier space images. We then employ the Gerchberg-Saxton phase retrieval algorithm\cite{gerchberg1994practical} to reconstruct the quantum state probability amplitudes and relative phases from the experimentally obtained real and Fourier space images. The experimentally obtained Fourier-space image is then compared with numerical simulations for the ideal case scenario of uniform coherent superposition. We observe a great similarity between these images, with both presenting similar interference patterns. Such a pattern is not presented by incoherent statistical mixtures, which leads us to conclude that the final state is indeed the W state. 

Compared with previous experiments \cite{grafe2014}, our approach  stands out for the on-demand nature of the quantum state generation, large operational bandwidth offered by the {Y-splitter-based architecture, and the better scalability and smaller circuit size offered by our state analysis protocol.

\begin{figure}
\centering
\includegraphics[width=1.0\linewidth]{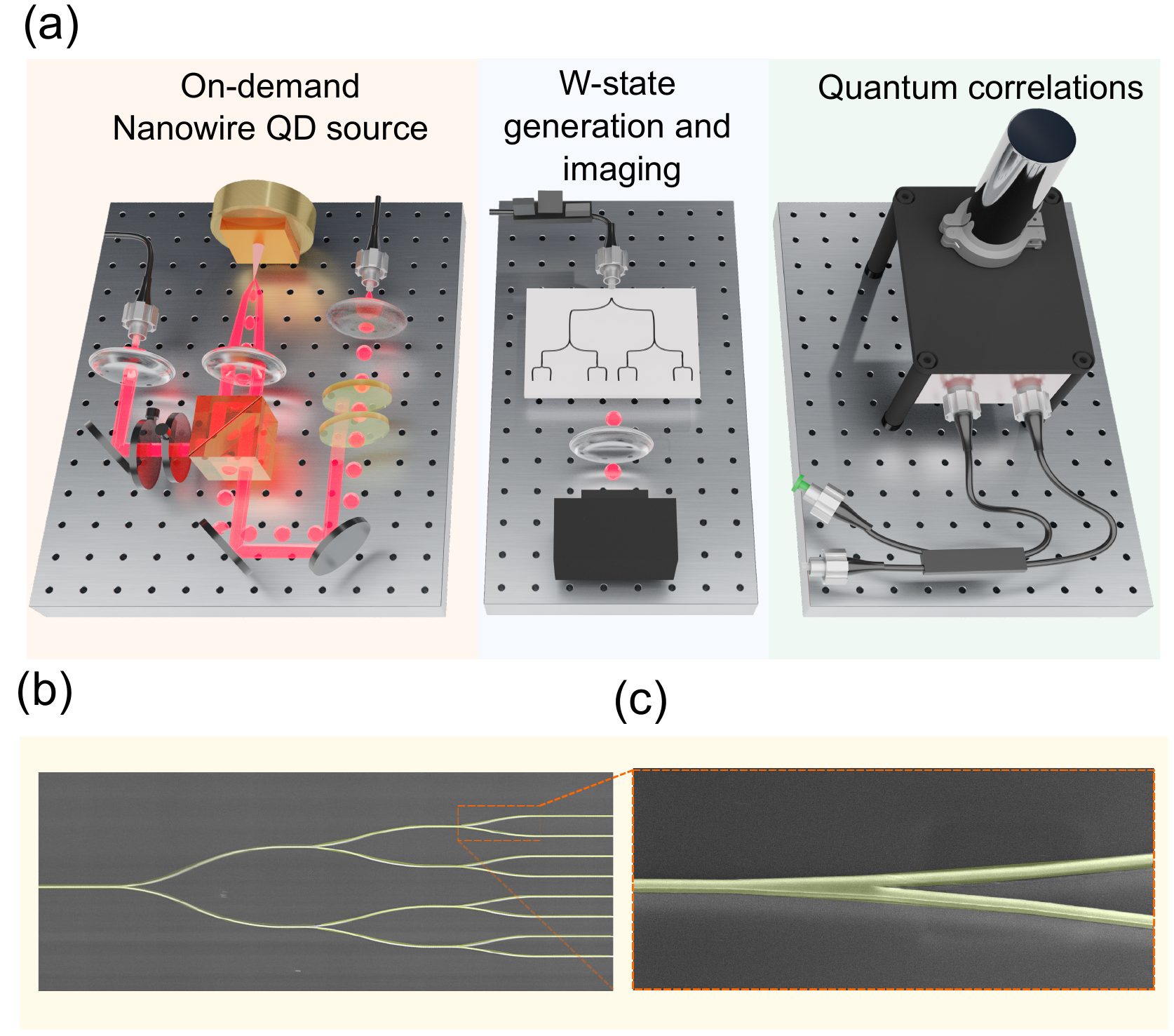}
\caption{\label{fig:figure1}\textbf{ (a) Schematic of the experimental setup.} The nanowire quantum dot (QD) is mounted in a closed-cycle cryogenic system operating at 4.2 K. The QD is excited using a tunable 320 MHz pulsed laser operating at a wavelength of 780 nm. The polarization of the laser is selected using a set of quarter-wave plates, a half-wave plate and a linear polarizer. Single photons from the QD are filtered using a long pass filter for laser rejection, and a cascade of tunable long-pass, short pass and band pass filters mounted on a rotating stage to select a single transmission from the S-shell of the QD. The charged exciton line $X^-$, shown in Fig.\ref{fig:figure2}(a), is then coupled to a single mode optical fiber. The single photons could be: (1) coupled to a fiber-based 50:50 beam-splitter with polarization controllers to a pair of superconducting nanowire single photon detectors for characterization of the second order correlation function. (2) coupled to the photonic-chip W state generator using a tapered optical fiber. The polarization of the single photons is set to the transverse electric (TE) field mode of the single mode photonic waveguide using a fiber-based polarization controller. The output of the chip is imaged using a 100X objective to a qCMOS Hamamatsu single-photon sensitive camera. The Fourier and real space images can be obtained by either adding or removing an additional optical lens before the camera. \textbf{SEM images of the fabricated devices.} (b) and (c) show the false-color SEM image of representative 8-mode W state device and a single Y-splitter, respectively (The size of the imaged device is different from the actual one used for the experiment. The actual Y-splitter array has a total length of 1mm and width of 26 $\mu$m at the output. An image of this aspect ratio would not be suitable to show the circuit details ).}
\label {fig1}
\end{figure}

We prepare the W state with an on-demand single photon source, as shown in the experimental setup Fig.~\ref{fig:figure1}(a). The on-demand single photon source consists of an InAsP quantum dot (QD) embedded in a wurtzite InP nanowire \cite{dalacu2012ultraclean,elshaari2017chip,zadeh2016deterministic,elshaari2018strain,gourgues2019controlled}, and further details on the corresponding nanowire growth process can be found in the \textbf{Supporting Information}. The nanowire quantum devices were maintained at 4.2 K in an \textit{attocube} closed-cycle cryogenic system. The single photon source was excited using 780 nm pulsed laser beam with a repetition rate of 320 MHz and an excitation power of 100 nW. A linear polarizer, a set of quarter-wave plates, and a half-wave plate are used to purify the laser's polarization. From among 100s of tested ultra-bright single photon sources, we selected the optimum emitter in terms of emission wavelength, brightness and emission linewidth. A long pass filter is used to reject laser light from single photons emitted by the nanowire quantum dot. A cascade of adjustable long-pass, short-pass and band-pass filters, placed on a rotating stage, are then used to select a single transition from the QD's S-shell. After coupling the single photons to an optical fiber, the photons can be either connected to a Hanbury Brown and Twiss effect (HBT) \cite{hanbury1997correlation, hanbury1993test} setup to measure the second order correlation function, or coupled, using a tapered optical fiber, to the photonic chip for W state generation. The output of the photonic chip is imaged using a qCMOS single photon sensitive camera by Hamamatsu. Real and Fourier space intensity images of the chip-output can projected to the camera using a combination of 100X objective and an optical lens. Fig.~\ref{fig:figure1}(b) and (c) show scanning electron microscope image of the W state device, and magnified image of a single Y-splitter.

\begin{figure}
\centering
\includegraphics[width=1.0\linewidth]{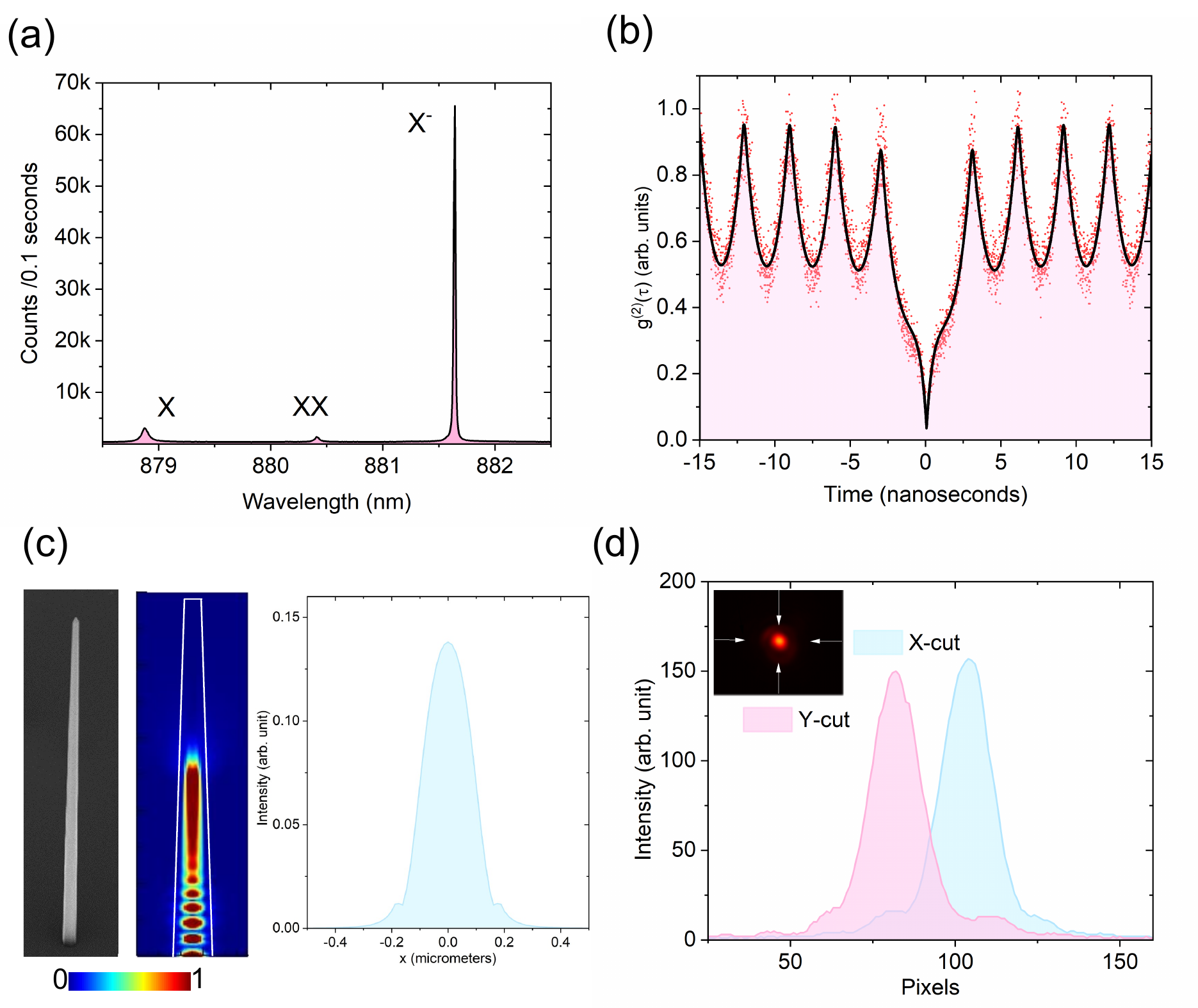}
\caption{\label{fig:figure2}\textbf{Nanowire-QD source}.  \textbf{(a) Emission spectrum of the nanowire QD}. At 100 nW laser excitation power, three emission lines are visible from the recombination of carriers in the S-shell of the QD. Using power and polarization series measurements, we identified the three emission lines to be the neutral exciton $X$ , the bi-exciton $XX$ , and the charged exciton (trion) $X^-$. We observed that the charged exciton $X^-$ is dominant under high energy laser excitation, consistent with previous measurements of similar nanowires. The nanowires exhibit excellent emission properties of small-linewidth and high brightness of several MHz detected on the superconducting nanowire single photon detectors. \textbf{(b) Single photon purity}. Second order correlation function $g^{(2)}(0)$ value at zero delay was measured to be 0.04. The black line shows fit to the experimental data points in red. The data is fitted with a series of correlation pulses at repetition rate of 320 MHz weighted with an antibunching term at zero-delay that describes the sub-Poissonoan statistics. \textbf{(c) Numerical simulation of nanowire QD mode profile}. 3D finite difference time domain simulations are performed to calculate the mode profile of the traveling waves in the nanowire. The source is a dipole oriented perpendicular to the growth direction, which is located 1.5 $\mu$m from the base of the nanowire. \textbf{(d) Measured emission-profile of the nanowire QD}. To facilitate the coupling to optical fibers with high efficiency, the emission profile of the single photons emitted from the nanowire was experimentally measured, following the numerical simulations. There is an excellent agreement with the numerical simulations, yielding a Gaussian-like profile thanks to the waveguiding effect provided by the core-shell structure of the nanowire QD. }
\label {fig2}
\end{figure}

Fig.~\ref{fig:figure2}(a) shows the emission spectrum of the nanowire quantum dot. The S-shell emission shows 3 types of particle-complexes, a neutral exciton, a bi-excitons and a trion, which were all verified using power and polarization series measurements. We used the brightest trion line at a wavelength of 881.7 nm, generating single photons with emission rates in the MHz range, measured using a superconducting single photon detector\cite{chang2023nanowire,esmaeil2021superconducting,esmaeil2020efficient}. To characterise the purity of the emitted single photons, we conducted zero delay second order correlation measurement $g^{(2)}(0)$ using a fiber-based HBT setup equipped with two superconducting single photon detectors. The system efficiencies of the two detectors are 80$\%$ and 66$\%$, with timing jitter of 18 picoseconds and 11 picoseconds, respectively, and dark counts of less than 10 Hz. At zero delay, the measured value of $g^{(2)}(0)$ is 0.04, well below the high order emission level, allowing us to operate in the single photon-limit of the Hilbert space, the results are shown in Fig.~\ref{fig:figure2}(b). The non-zero value for the $g^{(2)}(0)$ is due to re-excitation of the quantum dot within the lifetime of the photon emission, and possible contributions from other states within the filtered emission range. 

To determine the mode profile of the single photons emitted from the nanowire, 3D finite difference time domain simulations were performed, the results are shown in Fig.~\ref{fig:figure2}(c). The QD is simulated as a dipole located 1.5 $\mu$m out from the base of the nanowire, the dipole orientation is perpendicular to the growth direction. The waveguiding, provided by the core-shell design and the tapering of the nanowire, forms a circularly symmetric mode-profile that enhances coupling to single mode fibers. To verify the beam profile experimentally, and enhance the coupling efficiency of the single photons to the optical fiber, the emission profile of the single photons was measured as shown in Fig.~\ref{fig:figure2}(d). The results show excellent agreement with the numerical simulations, with a Gaussian-like emission. The mode profile of the QD emission was matched to a 780HP single mode fiber using a Schäfter+Kirchhoff fiber coupler with an adjustable aspherical lens to achieve a high coupling efficiency.

The fiber-coupled single photons are then injected to the W state photonic chip using a tapered optical fiber with a working distance of 13 $\mu$m and spot-size of 3 $\mu$m to maximize coupling to the transverse electric (TE) field mode of the photonic waveguides. The waveguides are made of silicon nitride deposited by Low Pressure Chemical Vapor Deposition (LPCVD) technique and then lithographically patterned to a width of 500 nm and a height of 250 nm, ensuring single mode operation at the nanowire quantum dot emission wavelength. The waveguides are coated with PMMA to ensure symmetric mode confinement. More details about the photonic chip fabrication can be found in the \textbf{Supporting Information}. The optical W state based on channel waveguides is characterized by a coherent distribution of a single photon over $N$ waveguides. The state is defined by
\begin{equation}\label{eq:5} 
|\mathrm{W}_N\rangle = \frac{1}{\sqrt{N}} \sum_{n=1}^N \exp(i\phi_n)a_n^\dag|\mathrm{0}\rangle,
\end{equation}
where $\phi_n$ is the arbitrary phase and $a_n^\dag$ is the Bosonic creation operator at each channel. In our experiment, the single photon W state generation occurs through coherent evolution of single photons through cascaded arrays of 3 sets of Y-branch 50-50 power splitters. In our circuit, every Y-branching was made to be precisely transversely symmetrical, providing an identical pathlength from input to output, regardless of the path. The emission lifetime of our single photon source is in the order 1 ns which is much longer than the pathlength corresponding to the physical dimensions of the chip. This ensures the presence of only a single photon in the chip at a time. In comparison to previously demonstrated methods employing directional couplers\cite{grafe2014, menotti2016generation} and evanescent coupling in waveguide arrays\cite{perez2013generating, swain2020single}, the Y-splitters based protocol is easier to design and scalable with larger operating bandwidth, limited by the single mode cut-off of the photonic waveguide.

The single photon state, $a_1^\dag|\mathrm{0}\rangle$ (where $a_1^\dag$ is the creation operator of the input waveguide), launched into the input waveguide is initially localized. Its state after evolution through the first y-splitter can be expressed as a 2-order W state \cite{ivanova2016using}
\begin{equation}
|\mathrm{W_2}\rangle=\frac{1}{\sqrt{2}}(b_1^\dag+b_2^\dag)|\mathrm{0}\rangle,  
\end{equation}
 where $b_1^\dag$ and $b_2^\dag$ are the creation operators at the outputs of the first Y branch. 

Similarly, the state after the second set of 2 Y-splitters reads 
\begin{equation}
|\mathrm{W_4}\rangle=\frac{1}{2}(c_1^\dag+c_2^\dag+c_3^\dag+c_4^\dag)|\mathrm{0}\rangle,
\end{equation}
and, finally, the final output state after the third set of 4 Y-splitters is the 8-order W state
\begin{equation}\label{eq:7}
|\mathrm{W_8}\rangle=\frac{1}{\sqrt{8}}(d_1^\dag+d_2^\dag+d_3^\dag+d_4^\dag+d_5^\dag+d_6^\dag+d_7^\dag+d_8^\dag)|\mathrm{0}\rangle.
\end{equation} 
Here, $c^\dag$ and $d^\dag$ are the creation operators for the second and third set of Y-branch outputs respectively. Therefore, as ideally a single photon is sent to the circuit, the final state produced will be an optical 8-order W state given by the above equation with equal relative phases [cf. Eq.~(\ref{eq:5})].

\begin{figure}
\centering
\includegraphics[width=1.0\linewidth]{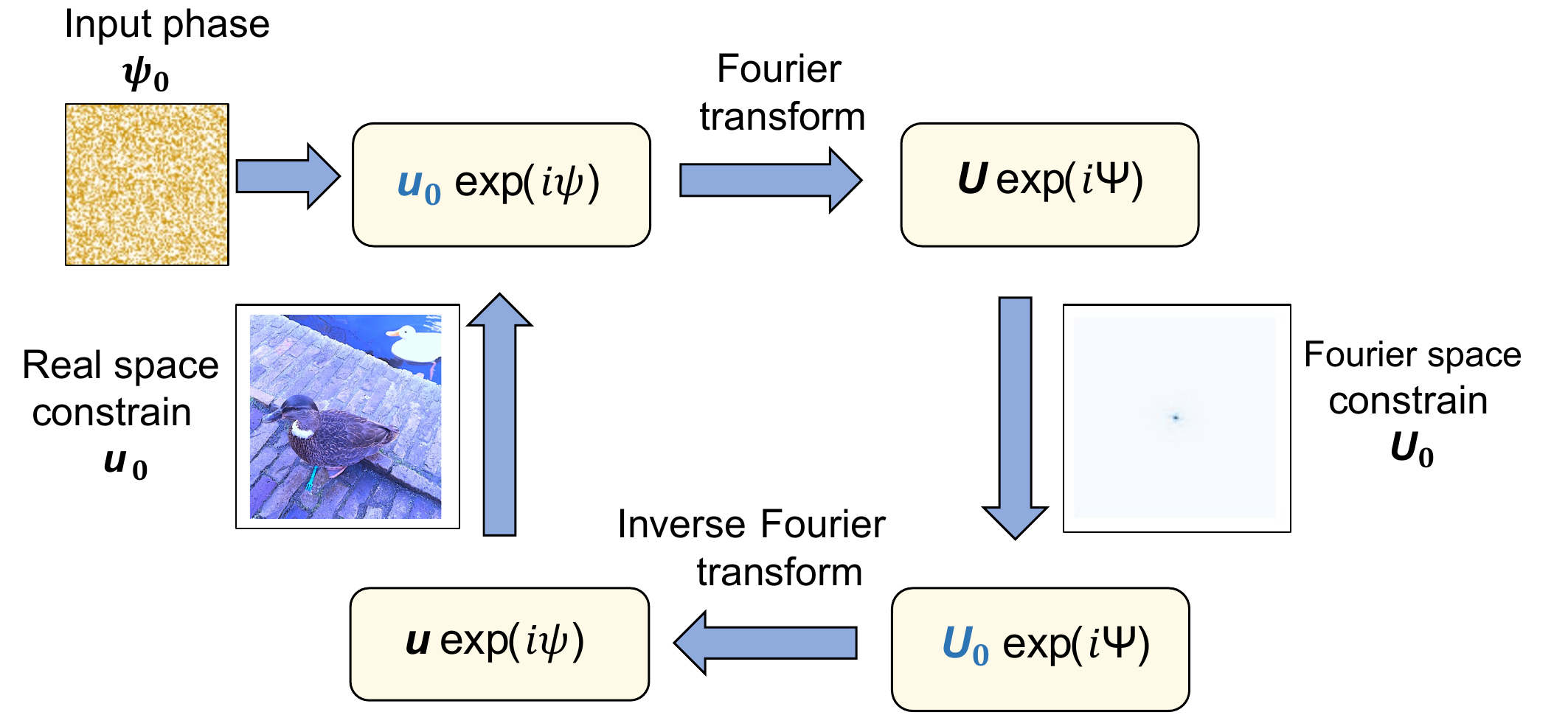}
\caption{\label{fig:figure3}\textbf{Work-flow diagram of Gerchberg-Saxton algorithm}. After the single photons are coupled to the photonic chip, intensity measurements of the real and Fourier space output-waveguides profiles are performed. The algorithms starts with the real space image initialized with a random phase distribution, serving as the basis for the optimization procedure. Then, while adhering to constrains imposed by the experimentally measured intensity distributions of the real and Fourier-space images, the Fourier and inverse Fourier transforms are carried out iteratively.  }
\label {fig3}
\end{figure}

After coupling the single photons to the chip input, the output facet of the chip was imaged using a qCMOS Hamamatsu camera. The Fourier and real space images can be obtained by either adding or removing an additional optical lens before the camera. In the experiment, we use the Gerchberg-Saxton algorithm  which was devised by crystallographers Ralph Gerchberg and Owen Saxton to deduce the phase distribution of electron beams in a transverse plane from the intensity distributions in two planes \cite{hanbury1997correlation, hanbury1993test}. A process flow diagram of the algorithm is shown in Fig.~\ref{fig:figure3}. The output phase distribution of our circuit can be reconstructed using this iterative phase retrieval process. The algorithm takes the 2D matrix corresponding to the real space amplitudes $u_0$ as the input and each point in the matrix is assigned an arbitrary phase value $\psi$. A Fourier transform operation on this matrix (with elements $u_0 \exp(i\psi)$) gives a Fourier space matrix $U\exp(i\Psi)$ which can in turn give a real space matrix with the application of inverse Fourier transform on it. Several iterations of this scheme is performed and each iteration yields a matrix with a set of either real space or Fourier space amplitudes and phases. After each transform operation, the amplitudes in the output matrix ($u$, $U$) are replaced by the amplitudes from the experimentally obtained real and Fourier images ($u_0$, $U_0$). These serve as the constraints in the algorithm. The phase values ($\psi$ in real space and $\Psi$ in Fourier space) are left unchanged and can evolve freely. The iterations continue until we get a convergence yielding real and Fourier space matrix amplitudes ($u$, $U$) very close to the experimentally observed ones ($u_0$, $U_0$). The phase that evolved freely, now converges to certain values which is equal to the actual relative phases of the real space image.

\begin{figure}
\centering
\includegraphics[width=1.0\linewidth]{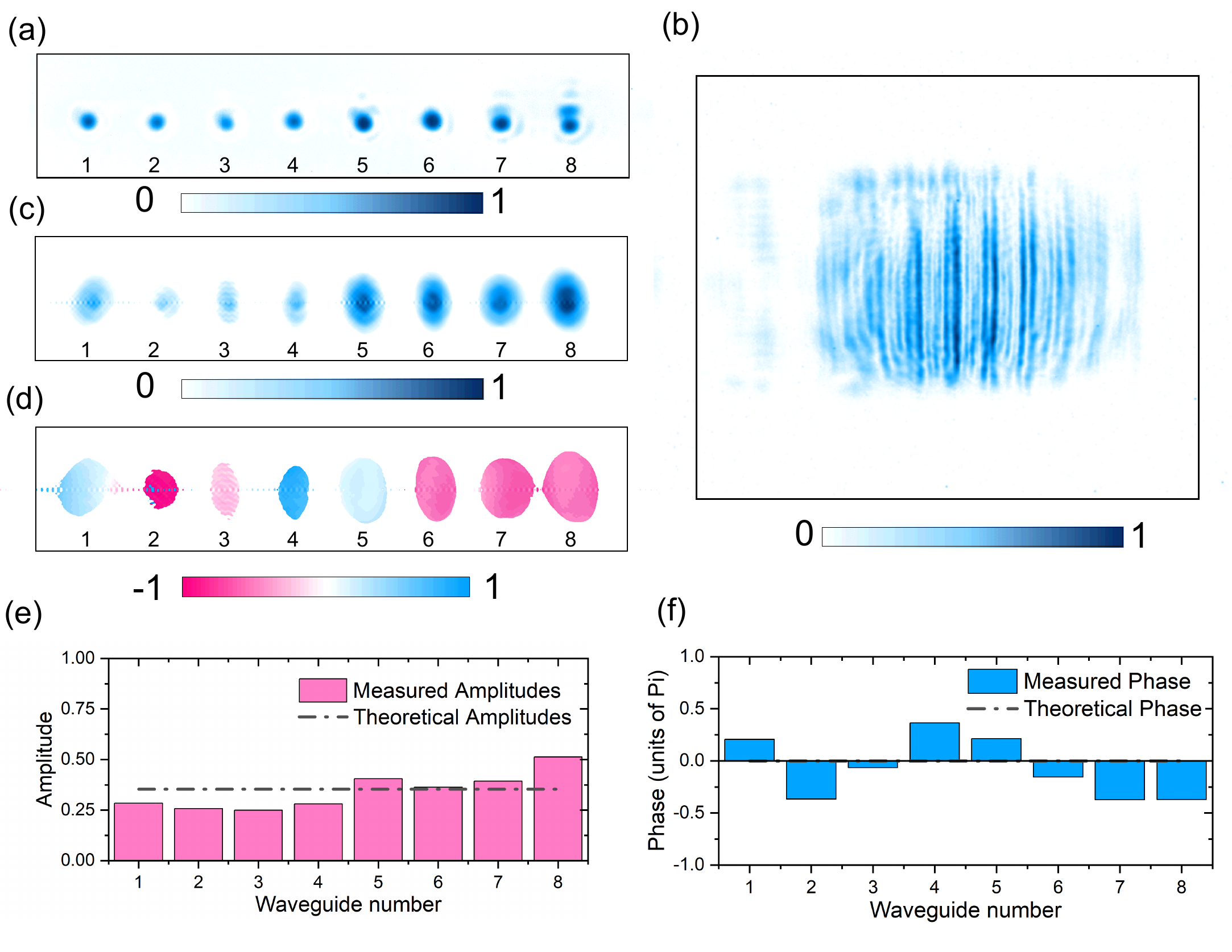}
\caption{\label{fig:figure4}\textbf{Quantum state reconstruction}. \textbf{(a) and (b), square-root of the real and Fourier space images}. The  real space image consists of 8 modes, with single photons coherently distributed between them. The Fourier space image emphasize this fact through a clear diffraction pattern created by the single photons. The interference pattern results from the coherent propagation of single photons through the chip, with no which-path information available. The images were recorded for an integration time of 10 minutes. \textbf{(c) and (d), reconstructed amplitude and phase profiles of the real space image}. The starting real and Fourier space images each contain 4 million pixels, the algorithm converged after 5000 interactions. There is a good agreement between the reconstructed and measured real space images, deviations in the reconstructed image are attributed to noise during data collection, which can be further reduced with enhanced coupling to the photonic chip The phase in (d) is measured in units of $\pi$. \textbf{(e) and (f), extracted amplitudes and phases of the W state}. The theoretical values of the amplitude ($1/\sqrt{8}$) and phase ($0$) are highlighted by a black-dotted line.  }
\label {fig4}
\end{figure}

Fig.~\ref{fig:figure4}(a) shows the real space amplitude (square-root of intensity) image of the output facet obtained using an exposure time of 10 minutes. Fig.~\ref{fig:figure4}(b) shows the Fourier space image obtained using the same experimental setup. There are 8 modes in the real space image corresponding to the 8-waveguides from the cascaded Y-splitters, and single photons are coherently distributed among them. This fact is highlighted in the Fourier space image by a distinct diffraction pattern produced by the individual photons ($g^{(2)}(0)=0.04$). The interference is the consequence of coherent single-photon superposition across the photonic chip without any knowledge of the which-path information. The real and Fourier space images in the experiment each had 4 million pixels, the Gerchberg-Saxton phase retrieval algorithm was run for 5000 iterations before convergence. The reconstructed amplitude and phase distributions of the W state from the experiment are shown in Fig.~\ref{fig:figure4}(c) and (d), respectively. The algorithm was able to successfully reconstruct the 8-mode real space image by directly taking the inverse Fourier transform of the reconstructed Fourier-space image. The degree of similarity of the reconstructed real space image with the experimentally obtained real space image is a measure of the accuracy of the derived phase values.  Fig.~\ref{fig:figure4}(e) and (f) show extracted amplitudes and phases of the experimentally measured on-demand W state generated by our photonic chip. We could observe good uniformity in the output probability amplitudes with a standard deviation of 0.085 around the mean value of 0.343 and a standard deviation of 0.086 around the ideal value of 0.354. The obtained phase values are also close to the ideal value of 0. The finite deviation from the ideal values in both cases is mainly due to the slight imperfections in the fabricated nanostructures, and background noise scattered in the cladding during image acquisition.

\begin{figure}
\centering
\includegraphics[width=1.0\linewidth]{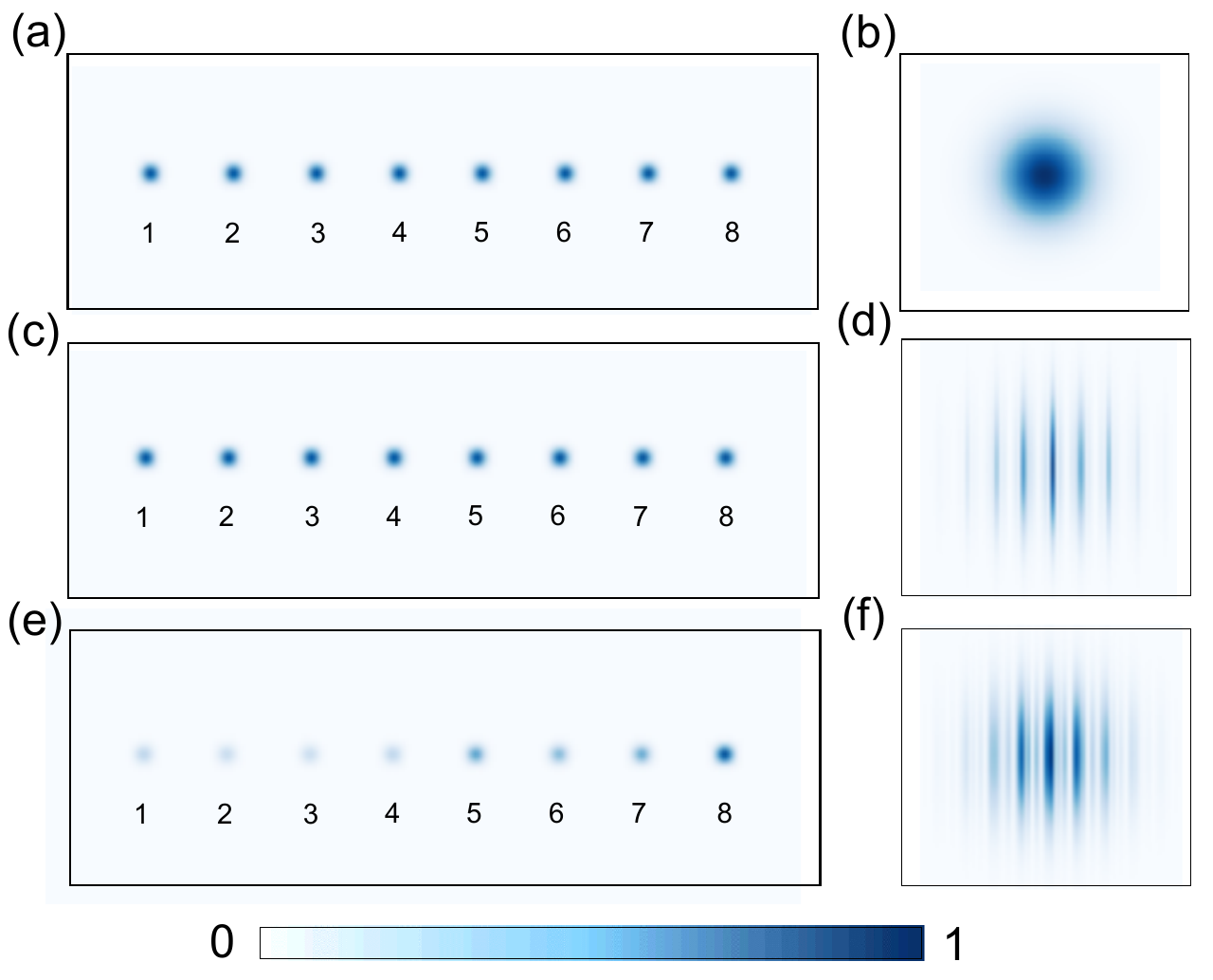}
\caption{\label{fig:figure5}\textbf{Verification of the multipartite coherent superposition}. \textbf{(a) and (b) real and Fourier space images of a mixed state}. 8-mode single photon mixed state is numerically generated, then its Fourier transform was computed. The interference pattern in the Fourier transform vanishes, yielding a Fourier transform profile corresponding to a single mode of the input.\textbf{(c) and (d) real and Fourier space images of pure W state}. The 8-mode single photon W state is numerically generated with equal amplitudes of ($1/\sqrt{8}$) and and a relative phase of ($0$) between different modes. The Fourier transform shows a clear interference as opposed to the mixed state case. \textbf{(e) and (f) real and Fourier space images experimentally measured W state}. Using the experimentally measured an phases, we numerically constructed the W state in the experiment and its Fourier transform. We clearly see an interference pattern in the Fourier transform, in close agreement with the measurements in  Fig.~\ref{fig:figure3}. This is in clear contrast to the theoretical results of a mixed state shown in Fig.~\ref{fig:figure4}(a).  }
\label {fig5}
\end{figure}

Traditionally, W states have been identified through state tomography and entanglement witnesses. Proper implementation of such techniques allows rigorous verification of the presence of multipartite entanglement or even reconstructing the obtained state. However, the complexity of implementing these techniques prohibits their application for quantum states involving a high number of qubits. So, it is desirable to find a scalable approach for higher-order W state verification. Here we propose a simple method based on image comparison techniques combined with reasonable assumptions about the experimental setup.

We start with some optical 8-order $W$ state [Eq. (5)]: $8^{-1/2}\sum_{n=1}^8 \exp(i\phi_n)a_n^\dagger |0\rangle$. The output image in the real space is composed of eight spots, each one corresponding to a term $\exp(i\phi_n)a_n^\dagger |0\rangle$. The Fourier space image, in turn, is obtained via 
\begin{equation}
\mathrm{FT}\Big(N^{-1/2}\sum_n \exp(i\phi_n)a_n^\dag|\mathrm{0}\rangle\Big),
\end{equation}
where $\mathrm{FT}(|\psi\rangle)$ stands for the Fourier transform of the image generated by $|\psi\rangle$. In the case of an ideal W state, $8^{-1/2}\sum_{n=1}^8 a_n^\dagger |0\rangle$, the real space image corresponds to 8 identical Gaussian spots and the Fourier space image resembles that of $n$ slits experiment with a characteristic interference pattern arising from the coherent superposition between different distinguishable states. In contrast to this, if the state undergoes complete decoherence its real space image remains the same but the interference pattern in the Fourier space image disappears since its Fourier transform now reads
\begin{equation}
N^{-1/2}\sum_n \mathrm{FT}\Big(\exp(i\phi_n)a_n^\dag|\mathrm{0}\rangle\Big).
\end{equation}
In Fig.\ref{fig:figure5} we compare the ideal W state images, its fully decohered version (mixed state) and the experimentally obtained image. Visually, we can clearly see that the result obtained experimentally resembles the simulation for the ideal W state, being in strong contrast with the mixed state. This is indeed confirmed since the images are more than 90$\%$ similar when compared via the Structural Similarity Index Measure (SSIM) \cite{wang2004image}. Furthermore, by computing the correlation between these images we can estimate the overlap between the ideal W state and the experimentally produced state, from which we get $83.1\%$. We now employ an entanglement witness of the form $\mathcal{W}_{\alpha\beta\gamma}=\alpha \mathcal{P}_0+\beta \mathcal{P}_1+\gamma \mathcal{P}_2-|W_8\rangle\langle W_8|$ ($\mathcal{P}_i$ is the projection onto the subspace with $i$ excitations), under different assumptions. By assuming that the produced state is a convex mixture of the ideal W state and its fully decohered version, i.e., $p|W_8\rangle\langle W_8|+(1-p)\mathcal{P}_1/8$, we get $p>80\%$. Employing the methods described in Ref. \cite{guhne2009} and briefly reviewed in \textbf{Supporting Information} we numerically found $\mathcal{W}_{\alpha\beta\gamma}$ witnessing the entanglement of the generated state. Moreover, the value of the second order correlation function $g^{(2)}(0)$ at zero delay gives an upper bound on the probability of having more than one photon on the chip. It was experimentally measured [Fig. 2 (b)] to be 0.04. Considering multiple photon generation as another possible source of noise in generated W state, the final state would have the form $(1-q)[p|W_8\rangle\langle W_8|+(1-p)\mathcal{P}_1/8]+q\mathcal{P}_2/28$ if we neglect contributions from subspaces with more than two excitations. These states, with the value of $q$ upper bounded by 0.04, can also have their entanglement witnessed by $\mathcal{W}_{\alpha\beta\gamma}$.

We have demonstrated a scalable on-demand scheme for high-order on-chip single photon W state generation. The on-demand nature of our protocol, using nanowire QDs and silicon nitride hybrid system, facilitates the scope of its integrability into other hybrid quantum systems\cite{elshaari2020hybrid,moody20222022}, with potential bit-rates in the GHz range, limited only by the lifetime of the quantum emitter. Our circuit based on Y-splitters use no resonant or interference effects, thus delivering large operating bandwidth. The ease of fabrication and the Fourier-space imaging based verification of superposition and coherence, makes our approach scalable to higher order W states. Through experimental measurements and theoretical modeling, we showed strong evidence to verify that our output state is a
multipartite coherent superposition, as opposed to a mixed state in which interferences  between different channels vanish. Our findings pave the way for future developments in image processing methods and Fourier space analysis for characterizing multi-partite entangled quantum systems. Multipartite entanglement provides a great deal of room for phenomena that are not available in systems with just two subsystems, which makes it an active field of research, both from fundamental science and applications point of view\cite{lombardi2002teleportation,qiao2021multistage,das2022optimal}. Our results introduce a quantifiable visual approach to experimentally validate multi-partite entanglement, which can be of paramount importance for the experimental advancement of multi-particle quantum-information processing protocols.

\begin{acknowledgments}
 A. W. E acknowledges support from the Knut and Alice Wallenberg (KAW) Foundation through the Wallenberg Centre for Quantum Technology (WACQT), Swedish Research Council (VR) Starting, and Vinnova quantum kick-start project 2021. S. S. acknowledges support from VR Starting. V. Z. acknowledges support from the KAW and VR. S. S. acknowledges support from the Swedish Research Council (Starting Grant No. 2019-04821) and from the Göran Gustafsson Foundation. L. S. acknowledges support from the House of Young Talents of the University of Siegen.
 O. G. acknowledges support from the Deutsche Forschungsgemeinschaft (DFG, German  Research Foundation, project numbers 447948357 and 440958198), the Sino-German     Center for Research Promotion (Project M-0294), the ERC (Consolidator Grant 683107/TempoQ), and the German Ministry of Education and Research (Project QuKuK, BMBF Grant No. 16KIS1618K). The authors acknowledge support from Quantum Design for using AFSEM, a correlative AFM and SEM system, to characterize the fabrication process of our waveguides.
\end{acknowledgments}

\clearpage
\onecolumngrid

\subsection*{\textbf{Supplementary Information}}

\subsection*{Quantum entanglement of W-states}

\noindent Here, we briefly review quantum entanglement theory with a focus on W states. We restrict ourselves to two-level quantum systems (qubits). As usual, we fix a ``computational basis" defined by two orthogonal states: $|0\rangle$ and $|1\rangle$. To start with, we consider the simplest bipartite quantum system presenting entanglement, a pair of qubits A and B. A given pure state $|\psi\rangle_{\rm AB}$ is entangled if it cannot be written as a product, i.e., $|\psi\rangle_{\rm AB}\neq |\phi\rangle_{\rm A}\otimes |\tau\rangle_{\rm B}$ for all local states $|\phi\rangle_{\rm A}$ 
and $|\tau\rangle_{\rm B}$.

For two quantum states, one can ask more generally, whether one state can be transformed into the other via local operations and classical communication (LOCC) \cite{chitambar2014}. For pure bipartite
states this can be solved \cite{nielsen1999}, but in the
general case this is a hard and open question 
\cite{vicente2013}.

A slight, but significant generalization of LOCC is that of 
stochastic local operations and classical communication (SLOCC). 
These are LOCC transformation on a single copy of a state, but 
without imposing that the target state has to be achieved with certainty. In that case, two states are equivalent if each 
one of them can be converted into the other with a non-zero probability and vice versa. For two qubits there is only one equivalence class represented by the Bell state:
\begin{equation}\label{eq:1}
    |\Phi^+\rangle=\frac{1}{\sqrt{2}}(|00\rangle+|11\rangle).
\end{equation}
Any entangled state can be obtained via LOCC from the Bell state, 
and conversely, any entangled state can be transformed into the 
Bell state with nonzero probability via SLOCC.
This is one of the many facts that justify the designation of ``maximally entangled" for the Bell state.

When considering a system of three or more qubits, the situation becomes much more complex. For three qubits A, B and C, for example, a pure state $|\psi\rangle_{\rm ABC}$ may be written as a product in two different ways: total separability, when the state is written as a product of three local states, $|\psi\rangle_{\rm ABC}=|\alpha\rangle_{\rm A}\otimes |\beta\rangle_{\rm B}\otimes |\gamma\rangle_{\rm C}$; and biseparability if $|\psi\rangle_{\rm ABC}=|\phi\rangle_{\rm A}\otimes |\tau\rangle_{\rm BC}$, $|\psi\rangle_{\rm ABC}=|\phi\rangle_{\rm AB}\otimes |\tau\rangle_{\rm C}$ or $|\psi\rangle_{\rm ABC}=|\phi\rangle_{\rm B}\otimes |\tau\rangle_{\rm CA}$. If $|\psi\rangle_{\rm ABC}$ is neither fully separable nor biseparable then it is genuine multipartite entangled \cite{guhne2009}. Furthermore, pure genuine multipartite entangled states 
can be entangled in two inequivalent ways \cite{dur2000}, 
i.e., there exist two classes of states which cannot be transformed into another by SLOCC, in contrast to two qubits. The representatives
of these entanglement classes are the Greenberger-Horne-Zeilinger
(GHZ) state,
\begin{equation}\label{eq:2}
    |\mathrm{GHZ}\rangle=\frac{1}{\sqrt{2}}(|000\rangle+|111\rangle),
\end{equation}
and the W state
\begin{equation}\label{eq:3}
    |\mathrm{W}\rangle=\frac{1}{\sqrt{3}}(|100\rangle+|010\rangle+|001\rangle).
\end{equation}

The GHZ and W states are of central importance in quantum information science \cite{guhne2009,horodecki2009}. Both can lead to violations 
of Bell inequalities \cite{cabello2002,pan2000experimental}, with the GHZ state violating the famous Mermin inequality \cite{mermin1990} maximally and leading to the 
GHZ argument \cite{greenberger1990}. Contrary to that, the entanglement in the W state is robust against particle loss and the state is maximally entangled according
to the geometric measure of entanglement \cite{wei2003,steinberg2022}.

Both states can be generalized to systems with many qubits. However, for systems with more than three qubits, there are infinitely many equivalence classes via SLOCC, which makes characterization much more complex. The generalization of Eq. (\ref{eq:3}) for $N$ qubits reads
\begin{equation}\label{eq:4}
    |\mathrm{W}_N\rangle=\frac{1}{\sqrt{N}}(|100\dots 0\rangle+|010\dots 0\rangle+\dots +|000\dots 1\rangle). 
\end{equation}
This state presents a variety of properties that make it unique in the set of pure states of many qubits system. The first of these properties is that, although it generally does not lead to the maximum violation of the better-known Bell inequalities in contrast to the GHZ state, the W state is much more robust against particle loss \cite{guhne2009}, making W state a good candidate to encode quantum information. In fact, the marginal states of GHZ (\ref{eq:2}) are separable while the W state (\ref{eq:3}) is the state with maximum possible bipartite entanglement in the reduced two-qubits state. Last but not least, W states are high-dimensional quantum states that exhibit a high degree of entanglement and whose experimental generation can be implemented robustly and much less demanding than other quantum states (e.g., GHZ states). So, on-demand generation of W states is a valuable tool for quantum technologies since such states are highly entangled and quite robust against harmful effects of the surrounding environment.

\subsection*{Photonic circuit fabrication}

To design the waveguide, ellipsometry measurements to characterise the height of the silicon nitride and its refractive index were performed. The simulated mode profiles based on the experimental measurements are shown in Fig.~\ref{fig:SFig1}(a), for the transverse electric TE (a) and transverse magnetic TM (b) modes. The TM mode is weakly localized, with an effective index of 1.588, compared to the TE mode which has an effective index of 1.643. The simulations are performed at the emission wavelength of the trion line of the QD. 

\begin{figure}[htbp]
\centering
\includegraphics[width=0.6\linewidth]{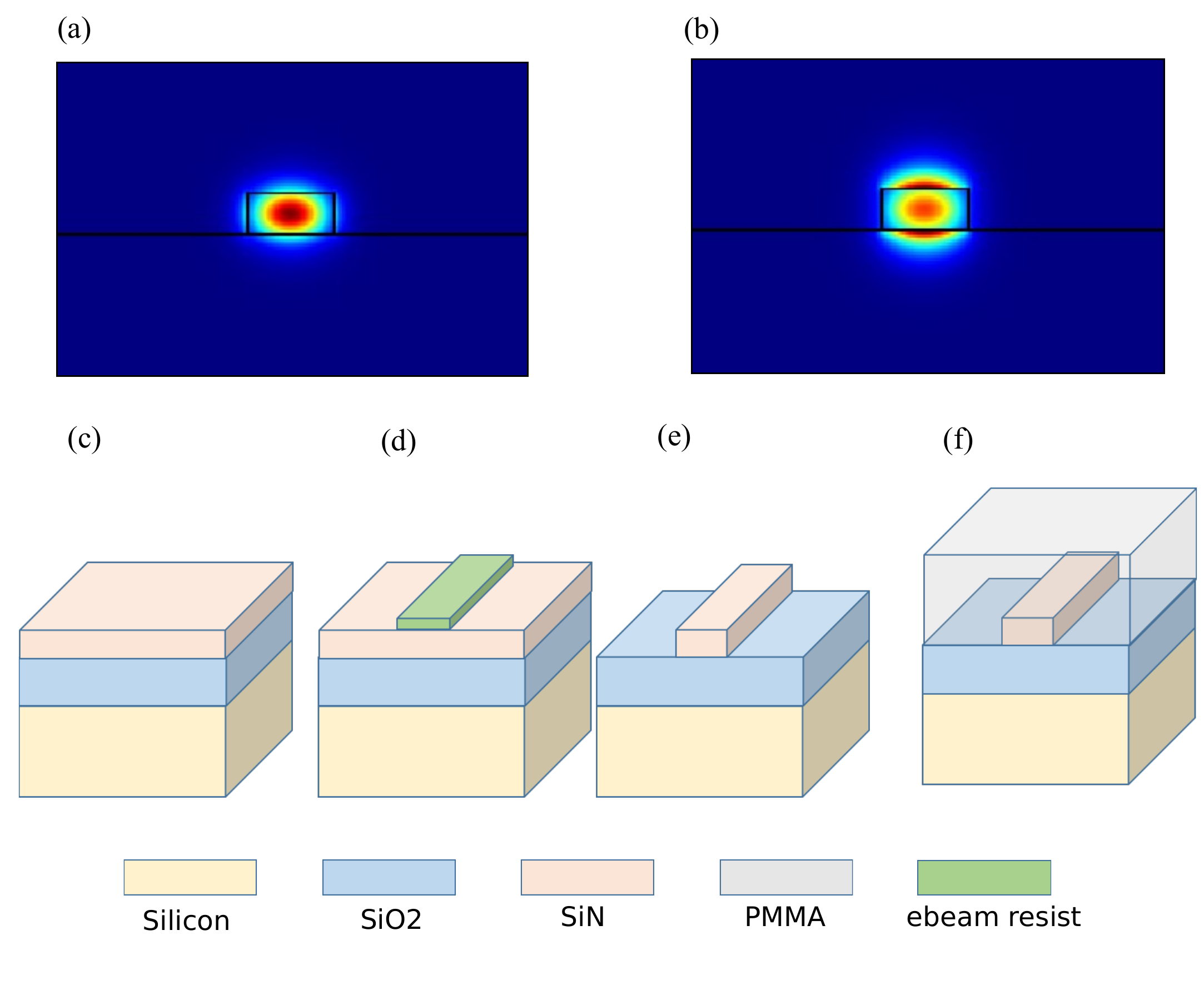}
\caption{\label{fig:SFig1}\textbf{Mode profiles.} (a) and (b) show Horizontal and vertical electric field components for the fundamental TE and TM modes, respectively. \textbf{W state chip fabrication process}. (c) Substrate consisting of a silicon wafer, 3.3 $\mu$m of silicon oxide and 250 $nm$ of silicon nitride. (d) electron beam lithography to define the pattern. (e) Reactive ion etching to form the circuit in the silion nitride layer. (f) top cladding of the chip with PMMA for symmetric mode confinement. }
\label {fig1}
\end{figure}

The substrate consists of a 500 $\mu$m thick silicon wafer capped with 3.3 $\mu$m of thermal oxide, and 250 $nm$ LPCVD silicon nitride prepared by Rogue Valley Microdevices. An adhesion promoter, AR 300-80 by ALLRESIST, is spin-coated on the substrate at 3000 RPM. The substrate is then soft-baked for 90 seconds at 90 $^{\circ}$C. Negative resist m-aN 2403 by Microresist technology is spin-coated at 3000 RPM and baked at 90 $^{\circ}$C for 1 minute, yielding an approximate resist thickness of 300 $nm$. The electron beam lithography dose assignment in the CAD is proximity-corrected using commercial software package BEAMER. The cad is exposed using 50 keV  electron-beam lithography Voyager system developed by Raith nanofabrication. The waveguide width was designed to be 600 nm. After exposure, the chip is developed in ma-D 525, an  aqueous-alkaline based developer, supplied by  Microresist technology, then the chip is rinsed in DI water. The waveguides are etched in PlasmaPro 100 Cobra ICP etching System using SF6 based chemistry. After etching, 950K A8 PMMA resist was spin-coated at 1000 RPM and baked at 150 $^{\circ}$C for 5 minutes. The refractive index of PMMA at 885 nm, the emission wavelength range of the S-shell transitions in the QD, is closely matched to the bottom oxide cladding. This provides symmetric mode confinement of the single photons in the silicon nitride waveguide. Finally, the chip is cleaved through a crystallographic direction of the silicon wafer, providing a smooth chip-facet for coupling light into the waveguides using a tapered optical fiber. The fabrication steps are depicted in Fig.~\ref{fig:SFig1}(d) to (f).

\begin{figure}[htbp]
\centering
\includegraphics[width=0.6\linewidth]{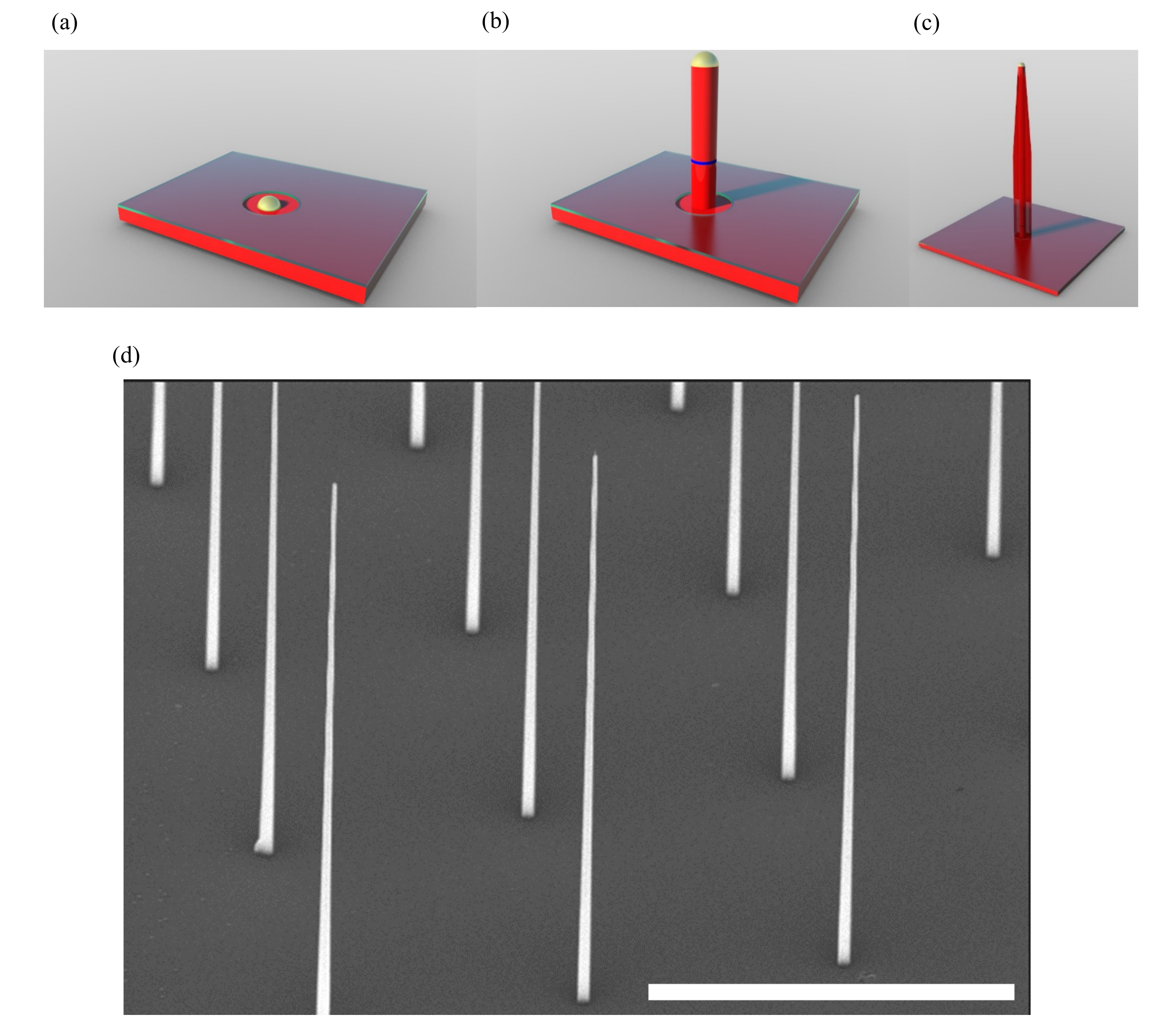}
\caption{\label{fig:SFig2}\textbf{Growth process of the nanowire quantum dot source.} (a) The process starts with a deterministically placed gold droplet to mediate the nanowire growth. (b) and (c) show the growth processes for the core, QD and cladding by controlling the chemistry and temperature, as highlighted in the text. \textbf{SEM of nanowires array}. (d) The scale bar has a length of 10\,$\mu$m. The controlled spacing between the nanowires allows for addressing a single nanowire-quantum dot using the laser.}
\label {fig2}
\end{figure}

\subsection*{Nanowire QD growth}
Chemical beam epitaxy using Trimethylindium (TMI), phosphine (PH$_3$) and arsine (AsH$_3$) as sources of In, P and As, respectively, was used to grow the wurtzite nanowires for this study. We us a selective-area vapour-liquid-solid growth technique described in detail in Refs.~\cite{Dalacu_APL2011,dalacu2012ultraclean,Laferriere_SR2022}. On a (111)B InP substrate we deposit a 20\,nm thick SiO$_2$ mask. Using electron-beam lithography, HF wet-etching and metal lift-off we produce patterned substrates consisting of gold droplets in the centres of holes in the SiO$_2$ mask. On this substrate we first grow InP nanowire cores which have a diameter of 20\,nm, set by the droplet size. In these cores we incorporate InAsP quantum dots $\sim 5$\,nm thick and having the same diameter as the core. We then clad the core with an InP shell to produce a photonic nanowire having a base diameter of 250\,nm which tapers to 100\,nm over the $\sim 15\,\mu$m length of the nanowire. The cladding is produced by adjusting the growth conditions from that used to grow the core, in particular, increasing the growth temperature from 435$^\circ$ to 450$^\circ$ and increasing the V/III ratio. The growth process is shown in Fig.~\ref{fig:SFig2}(a)-(c). A scanning electron microscope image of an array of deterministically fabricated nanowire quantum dots is shown in Fig.~\ref{fig:SFig2}(d).

\section*{Fourier-space images of W states}
The quantum interference between different channels in our system, which is revealed by the Fourier-space image, resembles that of $n$-slit experiment. To reveal this similarity, we constructed different quantum W states and computed their Fourier transform as shown in Fig.~\ref{fig:SFig3}. We selected W states of the following orders:
\begin{equation}\label{eq:1}
\begin{aligned}
\begin{split}
 &|\mathrm{W}\rangle=\frac{1}{\sqrt{1}}(|1\rangle) \\ 
 &|\mathrm{W}\rangle=\frac{1}{\sqrt{2}}(|10\rangle+|01\rangle) \\ 
 &|\mathrm{W}\rangle=\frac{1}{\sqrt{4}}(|1000\rangle+|0100\rangle+|0010\rangle+|0001\rangle)\\ 
 &|\mathrm{W}\rangle=\frac{1}{\sqrt{6}}(|100000\rangle+|010000\rangle+|001000\rangle+|000100\rangle+|000010\rangle+|000001\rangle)\\
 &|\mathrm{W}\rangle=\frac{1}{\sqrt{8}}(|10000000\rangle+|01000000\rangle+|00100000\rangle+|00010000\rangle+|00001000\rangle+|00000100\rangle\\
 &+|00000010\rangle+|00000001\rangle).\\
\end{split}
\end{aligned}
\end{equation}
In the trivial case of a single Gaussian beam input $g (x,y)$, W state of the 1$^{st}$ order, the far-field diffraction is the Fourier transform $(FT)$ of the input mode. The Fourier transform is a scaled Gaussian function, but in the spatial frequency space ($f_{x},f_{y}$).  The results are shown in Fig.~\ref{fig:SFig3}(a) and (b) and described
by
\begin{equation}\label{eq:2}
\begin{split}
\mbox{Interference pattern} = FT [g(x,y))] = G(f_{x},f_{y}).
\end{split}
\end{equation}

The situation becomes more interesting when more quantum channels are involved. For example, in the 2$^{nd}$ order W state, we can use the the translation property of the Fourier transform. If we assume that the two input modes are located at distances $\pm d$ from zero in the $x-$direction, the interference pattern can be written as
\begin{equation}\label{eq:3}
\begin{split}
\mbox{Interference pattern} = FT [g(x-d,y)+g(x+d,y))] = G(f_{x},f_{y})[e^{j2 \pi f_{x}d}+e^{-j2\pi f_{x}d}]
\end{split}
\end{equation}
The intensity of the interference pattern is simply a Gaussian function modulated by a squared cosine function. The input mode profile and the diffraction pattern intensity profile are shown in  Fig.~\ref{fig:SFig3}(c) and (d) and given by
\begin{equation}\label{eq:3}
\begin{split}
\mbox{Intensity profile} \sim G(f_{x},f_{y}) \cdot \cos^2(2\pi f_{x}d).
\end{split}
\end{equation}
As the number of modes is increased, we can write the diffraction pattern of the W state 
as
\begin{equation}\label{eq:4}
\begin{split}
\mbox{Intensity profile} \sim G(f_{x},f_{y}) \cdot  \frac{\sin^2(N \pi f_{x}d)}{\sin^2( \pi f_{x}d).}
\end{split}
\end{equation}
Here, $N$ is the number of interfering modes in the W state in Eq.~\ref{eq:1}. The position of the bright regions in the Fourier transform is preserved, following the same physics as in multi-slit diffraction. The results for the cases of 2, 4, 6, and 8 modes are shown in Fig.~\ref{fig:SFig3}(c) to (k). This is in a stark contrast to the mixed-state case, where the interference between different modes in the W state is lost, resulting in an incoherent mix of all the modes, with vanishing diffraction pattern.

 In our setup the Fourier transform can be computed by inserting a lens to project the W state output of the chip to the back-focal plane of the lens. The image in the back-focal plane of a positive lens, having a focal length $f$ and light of wavelength $\lambda$, 
 is given by 
\begin{equation}
\label{eq:6}
G\left( x\prime,  y\prime \right)=\exp \left\{j \pi \frac{x\prime^2+y\prime^2}{\lambda f}\left(1-\frac{z}{f}\right)\right\} \iint g(x, y) \exp \left\{-j 2 \pi \frac{x x\prime+y y\prime}{\lambda f}\right\} \mathrm{d} x \mathrm{~d} y,
\end{equation}
where \(x\prime\) \(y\prime\) are the transverse spatial coordinates after the lens at a distance \(z\). The output image is simply the 2-dimensional Fourier transform of the input, as shown by the integral over the input state in the second term. The integration is taken over the pupil function of the lens. The first term describes the spherical wave-front free-space propagation. 

When the camera is placed  exactly one focal distance behind the lens, the Fourier transform computed by the lens is exactly
\begin{equation}\label{eq:7}
G\left( x_{f},  y_{f}\right)=\iint g(x, y) \exp \left\{-j 2 \pi \frac{x x_{f}+y y_{f}}{\lambda f}\right\} \mathrm{d} x \mathrm{~d} y,
\end{equation}
where the spatial frequencies in the \(f_{x}\) and \(f_{y}\) direction are related to the spatial coordinates at the focal point   \(x_{f}\), \(y_{f}\) by \(f_{x}=x_{f}/ \lambda f\) and \(f_{y}=y_{f}/ \lambda f\). Moreover, the translation property of the Fourier transform can be understood in  our optical setup as shown in Fig.~\ref{fig:SFig4}. Different modes of the W state are focused to the back-focal plane, with each having a different path-length  corresponding to unique phase factor in the Fourier transform. This coherent locked phase between the modes enables the diffraction pattern we measure in the experiment between different single-photon channels.

\section*{Entanglement witnesses}

An entanglement witness is a self-adjoint operator $\mathcal{W}$ satisfying $\mathrm{tr}(\mathcal{W}\rho)\geq 0$ for all density operator representing a separable (i.e., non-entangled) quantum state. Thus, $\mathrm{tr}(\mathcal{W}\rho)$ being negative is a sufficient criterion to conclude that a given state $\rho$ is entangled. The construction of an adequate entanglement witness depends on some prior knowledge about the devices that produce such a quantum state. Here, in particular, we ideally produce the 8 order W state, $|W_8\rangle$. In that case, a good ansatz for entanglement witness is \cite{haffner2005scalable,guhne2009}
\begin{equation}
W_{\alpha\beta\gamma}=\alpha \mathcal{P}_0+\beta \mathcal{P}_1+\gamma\mathcal{P}_2-|W_8\rangle \langle W_8 |.
\end{equation}
Here $\mathcal{P}_i$ are projectors onto the subspaces with exactly $i$ excitations. We need to guarantee that $\mathcal{W}_{\alpha\beta\gamma}$ is actually an entangled witness. From the symmetry of the W state, it suffices to prove non-negativity for states $|a\rangle\otimes |b\rangle$, where $|a\rangle= a_0|00...00\rangle+a_1(|00...01\rangle + ...+ |10...00\rangle)$ and similarly for $|b\rangle$. Therefore, the problem of finding an entanglement witness for the produced state $\rho$ is read as
\begin{eqnarray}
\textrm{Find }&& \alpha\beta\gamma \nonumber\\
\textrm{Subject to }&& \langle ab|\mathcal{W}_{\alpha\beta\gamma}|ab \rangle \geq 0 \\
\textrm{and }&& \mathrm{tr}(\mathcal{W}_{\alpha\beta\gamma}\rho)<0.\nonumber
\end{eqnarray}
For $N=8$, the number of parameters allows this problem to be solved numerically. The states we consider in the main text have the form
\begin{equation}
\rho=(1-q)\left[p|W_8\rangle\langle W_8|+(1-p)\frac{\mathcal{P}_1}{8}\right]+q\frac{\mathcal{P}_2}{28}.
\end{equation}
The condition for $\rho$ being entangled then reads
\begin{equation}
(1-q)\left(\beta-\frac{7p+1}{8}\right)+\gamma q<0,
\end{equation}
given that $\mathcal{W}_{\alpha\beta\gamma}$ is a witness. For small values of $1-p$ and $q$, it is possible to find such a witness. In particular, we numerically verify it for $p>0.7$ and $q<0.2$.

\begin{figure}[htbp]
\centering
\includegraphics[width=0.6\linewidth]{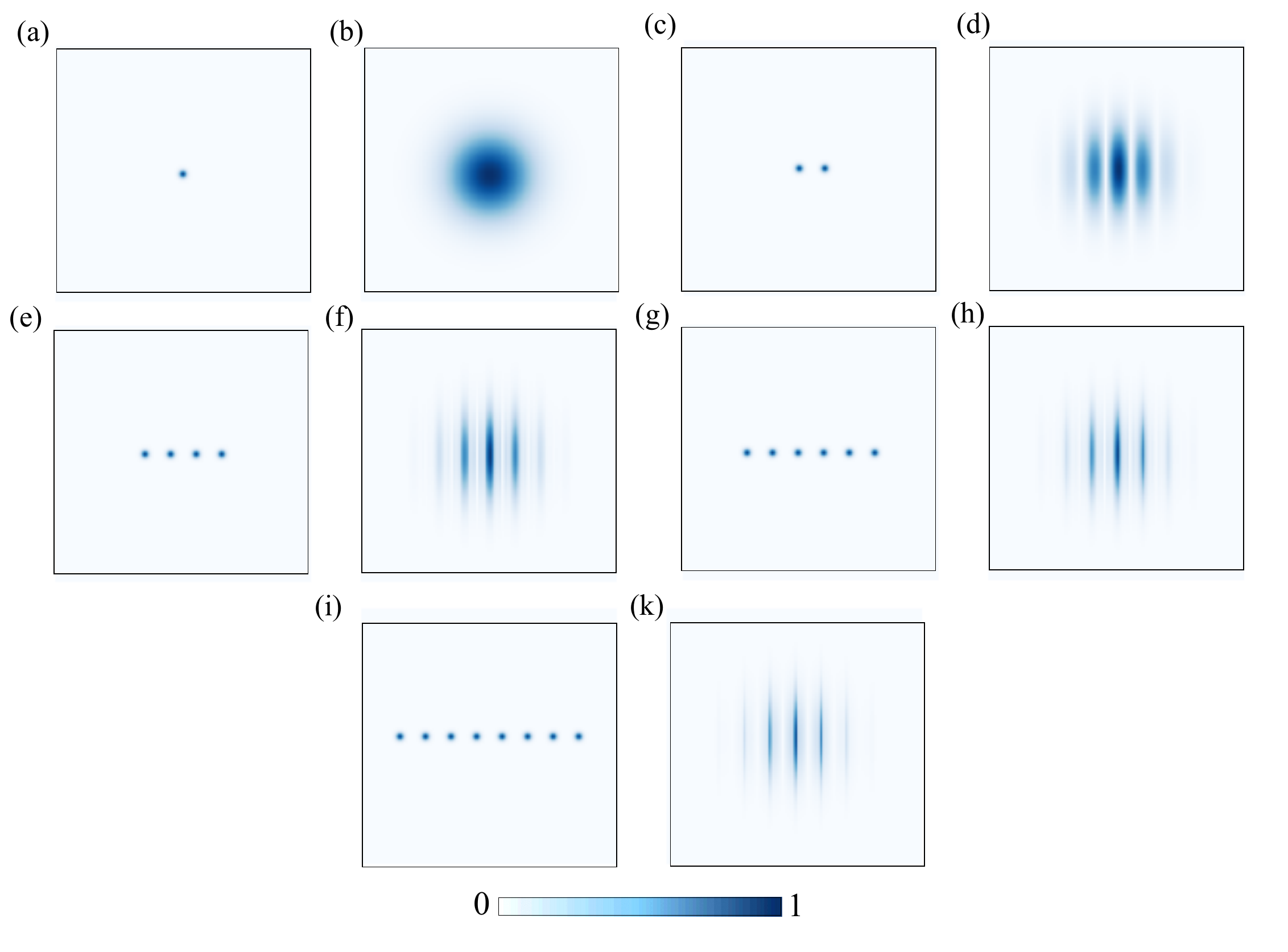}
\caption{\label{fig:SFig3}\textbf{W states in real and Fourier space}. (a), (c), (e), (g), (i) show real space images of W states of 1$^{st}$, 2$^{nd}$, 4$^{th}$, 6$^{th}$, and 8$^{th}$ order respectively. The corresponding Fourier transform images are shown in (b) ,(d) ,(f) ,(h) and (k). The simulated quantum states represent the ideal case of equal probability amplitudes in different channels, and selected relative phase difference of zero between them. }
\label {fig3}
\end{figure}

\begin{figure}[htbp]
\centering
\includegraphics[width=0.6\linewidth]{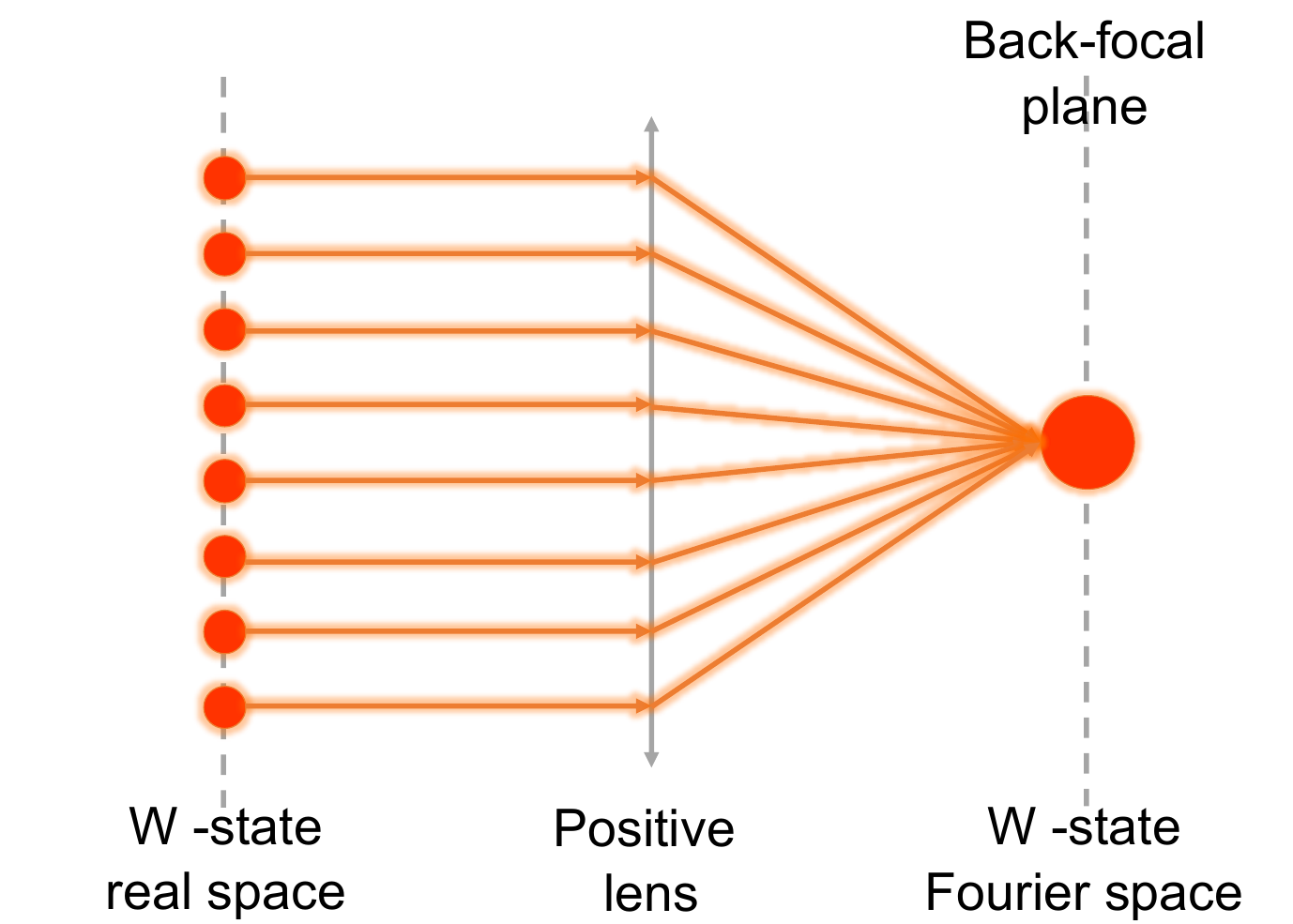}
\caption{\label{fig:SFig4}\textbf{Fourier transform using a positive lens}. Different modes in the W state are focused to the back-focal plane, with different path lengths corresponding to an overall phase factor of each mode. }
\label {fig4}
\end{figure}


\begin{thebibliography}{99}

    \bibitem{Dalacu_APL2011}
    D. Dalacu, K. Mnaymneh, X. Wu, J. Lapointe, G. C. Aers, P. J. Poole, R. L. Williams, Selective-area vapor-liquid-sold growth of tunable InAsP quantum dots in nanowires, \textit{Appl. Phys. Lett.} \textbf{98}, 251101 (2011).
    
    \bibitem{Laferriere_SR2022}
    P. Laferri{\`e}re, E. Yeung, I. Miron, D. B. Northeast, S. Haffouz, J. Lapointe, M. Korkusinski, P. J. Poole, R. L. Williams, D. Dalacu, Unity yield of deterministically positioned quantum dot single photon sources, \textit{Sci. Rep.} \textbf{12}, 6376 (2022).
    
    \bibitem{pearl2009causality}
    Judea Pearl, \textit{Causality}, Cambridge university press (2009).
    
    \bibitem{bell1964einstein}
    John S Bell, On the Einstein-Podolsky-Rosen paradox, \textit{Physics Physique Fizika}, \textbf{1}(3), 195 (1964).
    
    \bibitem{dicke1954coherence}
    Robert H Dicke, Coherence in spontaneous radiation processes, \textit{Physical Review}, \textbf{93}(1), 99 (1954).
    
    \bibitem{agrawal2006perfect}
    Pankaj Agrawal, Arun Pati, Perfect teleportation and superdense coding with W states, \textit{Physical Review A}, \textbf{74}(6), 062320 (2006).
    
    \bibitem{horodecki2009}
    Ryszard Horodecki, Pawe{\l} Horodecki, Micha{\l} Horodecki, Karol Horodecki, Quantum entanglement, \textit{Reviews of Modern Physics}, \textbf{81}(2), 865 (2009).
    
    \bibitem{bengtsson2017}
    Ingemar Bengtsson, Karol {\.Z}yczkowski, \textit{Geometry of quantum states: an introduction to quantum entanglement}, Cambridge university press (2017).
    
    \bibitem{guhne2009}
    Otfried G{\"u}hne, G{\'e}za T{\'o}th, Entanglement detection, \textit{Physics Reports}, \textbf{474}(1-6), 1--75 (2009).
    
    \bibitem{dur2000}
    Wolfgang D{\"u}r, Guifre Vidal, J Ignacio Cirac, Three qubits can be entangled in two inequivalent ways, \textit{Physical Review A}, \textbf{62}(6), 062314 (2000).
    
    \bibitem{grafe2014}
    Markus Gr{\"a}fe, Ren{\'e} Heilmann, Armando Perez-Leija, Robert Keil, Felix Dreisow, Matthias Heinrich, Hector Moya-Cessa, Stefan Nolte, Demetrios N Christodoulides, Alexander Szameit, On-chip generation of high-order single-photon W-states, \textit{Nature Photonics}, \textbf{8}(10), 791--795 (2014).
    
    \bibitem{cabello2002}
    Ad{\'a}n Cabello, Bell’s theorem with and without inequalities for the three-qubit Greenberger-Horne-Zeilinger and W states, \textit{Physical Review A}, \textbf{65}(3), 032108 (2002).
    
    \bibitem{barnea2015}
    Tomer Jack Barnea, Gilles P{\"u}tz, Jonatan Bohr Brask, Nicolas Brunner, Nicolas Gisin, Yeong-Cherng Liang, Nonlocality of W and Dicke states subject to losses, \textit{Physical Review A}, \textbf{91}(3), 032108 (2015).
    
    \bibitem{sohbi2015}
    Adel Sohbi, Isabelle Zaquine, Eleni Diamanti, Damian Markham, Decoherence effects on the nonlocality of symmetric states, \textit{Physical Review A}, \textbf{91}(2), 022101 (2015).
    
    \bibitem{murao1999}
    M Murao, D Jonathan, MB Plenio, V Vedral, Quantum telecloning and multiparticle entanglement, \textit{Physical Review A}, \textbf{59}(1), 156 (1999).
    
    \bibitem{shi2002}
    Bao-Sen Shi, Akihisa Tomita, Teleportation of an unknown state by W state, \textit{Physics Letters A}, \textbf{296}(4-5), 161--164 (2002).
    
    \bibitem{joo2003}
    Jaewoo Joo, Young-Jai Park, Sangchul Oh, Jaewan Kim, Quantum teleportation via a W state, \textit{New Journal of Physics}, \textbf{5}(1), 136 (2003).
    
    \bibitem{fang2019}
    B Fang, M Menotti, M Liscidini, JE Sipe, VO Lorenz, Three-photon discrete-energy-entangled w state in an optical fiber, \textit{Physical review letters}, \textbf{123}(7), 070508 (2019).
    
    
    \bibitem{erhard2020}
    Manuel Erhard, Mario Krenn, Anton Zeilinger, Advances in high-dimensional quantum entanglement, \textit{Nature Reviews Physics}, \textbf{2}(7), 365--381 (2020).
    
    \bibitem{hanbury1997correlation}
    R Hanbury Brown, RQ Twiss, Correlation between photons in two coherent beams of light, \textit{SPIE MILESTONE SERIES MS}, \textbf{139}, 93--95 (1997).
    
    \bibitem{hanbury1993test}
    R Hanbury Brown, RQ Twiss, A test of a new type of stellar interferometer on Sirius (from Nature 1956), \textit{SPIE MILESTONE SERIES MS}, \textbf{73}, 335--335 (1993).
    
    \bibitem{dalacu2012ultraclean}
    Dan Dalacu, Khaled Mnaymneh, Jean Lapointe, Xiaohua Wu, Philip J Poole, Gabriele Bulgarini, Val Zwiller, Michael E Reimer, Ultraclean emission from InAsP quantum dots in defect-free wurtzite InP nanowires, \textit{Nano letters}, \textbf{12}(11), 5919--5923 (2012).
    
    \bibitem{gerchberg1994practical}
    RW Gerchberg, WO Saxton, A practical algorithm for the determination of phase from image and diffraction plane pictures, \textit{SPIE milestone series MS}, \textbf{94}, 646--646 (1994).
    
    \bibitem{feng2019chip}
    Tianfeng Feng, Xiaoqian Zhang, Yuling Tian, Qin Feng, On-chip multiphoton entangled states by path identity, \textit{International Journal of Theoretical Physics}, \textbf{58}(11), 3726--3733 (2019).
    
    \bibitem{perez2013generating}
    Armando Perez-Leija, JC Hernandez-Herrejon, Hector Moya-Cessa, Alexander Szameit, Demetrios N Christodoulides, Generating photon-encoded W states in multiport waveguide-array systems, \textit{Physical Review A}, \textbf{87}(1), 013842 (2013).
    
    \bibitem{swain2020single}
    Manoranjan Swain, Amit Rai, M Karthick Selvan, Prasanta K Panigrahi, Single photon generation and non-locality of perfect W-state, \textit{Journal of Optics}, \textbf{22}(7), 075202 (2020).
    
    \bibitem{menotti2016generation}
    M Menotti, L Maccone, JE Sipe, M Liscidini, Generation of energy-entangled W states via parametric fluorescence in integrated devices, \textit{Physical Review A}, \textbf{94}(1), 013845 (2016).
    
    \bibitem{ivanova2016using}
    AE Ivanova, SA Chivilikhin, AV Gleim, Using of optical splitters in quantum random number generators, based on fluctuations of vacuum, In \textit{Journal of Physics: Conference Series}, \textbf{735}(1), IOP Publishing (2016).
    
    \bibitem{guo2002scheme}
    Guang-Can Guo, Yong-Sheng Zhang, Scheme for preparation of the W state via cavity quantum electrodynamics, \textit{Physical Review A}, \textbf{65}(5), 054302 (2002).
    
    \bibitem{zang2016}
    Xue-Ping Zang, Ming Yang, Fatih Ozaydin, Wei Song, Zhuo-Liang Cao, Deterministic generation of large scale atomic W states, \textit{Optics express}, \textbf{24}(11), 12293--12300 (2016).
    
    \bibitem{wang2001entanglement}
    Xiaoguang Wang, Entanglement in the quantum Heisenberg XY model, \textit{Physical Review A}, \textbf{64}(1), 012313 (2001).
    
    \bibitem{vandersypen2005nmr}
    Lieven MK Vandersypen, Isaac L Chuang, NMR techniques for quantum control and computation, \textit{Reviews of modern physics}, \textbf{76}(4), 1037 (2005).
    
    \bibitem{dogra2015}
    Shruti Dogra, Kavita Dorai, Experimental construction of generic three-qubit states and their reconstruction from two-party reduced states on an NMR quantum information processor, \textit{Physical Review A}, \textbf{91}(2), 022312 (2015).
    
    \bibitem{das2015}
    Debmalya Das, Shruti Dogra, Kavita Dorai, Experimental construction of a W superposition state and its equivalence to the Greenberger-Horne-Zeilinger state under local filtration, \textit{Physical Review A}, \textbf{92}(2), 022307 (2015).
    
    \bibitem{haas2014entangled}
    Florian Haas, J{\"u}rgen Volz, Roger Gehr, Jakob Reichel, J{\'e}r{\^o}me Est{\`e}ve, Entangled states of more than 40 atoms in an optical fiber cavity, \textit{Science}, \textbf{344}(6180), 180--183 (2014).
    
    \bibitem{hosten2016measurement}
    Onur Hosten, Nils J Engelsen, Rajiv Krishnakumar, Mark A Kasevich, Measurement noise 100 times lower than the quantum-projection limit using entangled atoms, \textit{Nature}, \textbf{529}(7587), 505--508 (2016).
    
    \bibitem{mcconnell2015entanglement}
    Robert McConnell, Hao Zhang, Jiazhong Hu, Senka {\'C}uk, Vladan Vuleti{\'c}, Entanglement with negative Wigner function of almost 3,000 atoms heralded by one photon, \textit{Nature}, \textbf{519}(7544), 439--442 (2015).
    
    \bibitem{frowis2017experimental}
    Florian Frwis, Peter C Strassmann, Alexey Tiranov, Corentin Gut, Jonathan Lavoie, Nicolas Brunner, F{\'e}lix Bussi{\`e}res, Mikael Afzelius, Nicolas Gisin, Experimental certification of millions of genuinely entangled atoms in a solid, \textit{Nature communications}, \textbf{8}(1), 1--6 (2017).
    
    \bibitem{pu2018experimental}
    Yunfei Pu, Yukai Wu, Nan Jiang, Wei Chang, Chang Li, Sheng Zhang, Luming Duan, Experimental entanglement of 25 individually accessible atomic quantum interfaces, \textit{Science advances}, \textbf{4}(4), eaar3931 (2018).
    
    \bibitem{li2021multipartite}
    Hang Li, Jian-Peng Dou, Xiao-Ling Pang, Chao-Ni Zhang, Zeng-Quan Yan, Tian-Huai Yang, Jun Gao, Jia-Ming Li, Xian-Min Jin, Multipartite entanglement of billions of motional atoms heralded by single photon, \textit{npj Quantum Information}, \textbf{7}(1), 1--9 (2021).
    
    \bibitem{roos2004}
    Christian F Roos, Mark Riebe, Hartmut Haffner, Wolfgang Hansel, Jan Benhelm, Gavin PT Lancaster, Christoph Becher, Ferdinand Schmidt-Kaler, Rainer Blatt, Control and measurement of three-qubit entangled states, \textit{Science}, \textbf{304}(5676), 1478--1480 (2004).
    
    \bibitem{haffner2005scalable}
    Hartmut Häffner, Wolfgang Hänsel, CF Roos, Jan Benhelm, D Chek-al-Kar, M Chwalla, T Körber, UD Rapol, M Riebe, PO Schmidt, Scalable multiparticle entanglement of trapped ions, \textit{Nature}, \textbf{438}(7068), 643--646 (2005).
    
    \bibitem{papp2009characterization}
    Scott B Papp, Kyung Soo Choi, Hui Deng, Pavel Lougovski, SJ Van Enk, HJ Kimble, Characterization of multipartite entanglement for one photon shared among four optical modes, \textit{Science}, \textbf{324}(5928), 764--768 (2009).
    
    \bibitem{choi2010}
    KS Choi, A Goban, SB Papp, SJ Van Enk, HJ Kimble, Entanglement of spin waves among four quantum memories, \textit{Nature}, \textbf{468}(7322), 412--416 (2010).
    
    \bibitem{elshaari2020hybrid}
    Ali W Elshaari, Wolfram Pernice, Kartik Srinivasan, Oliver Benson, Val Zwiller, Hybrid integrated quantum photonic circuits, \textit{Nature Photonics}, \textbf{14}(5), 285--298 (2020).
    
    \bibitem{elshaari2017chip}
    Ali W Elshaari, Iman Esmaeil Zadeh, Andreas Fognini, Michael E Reimer, Dan Dalacu, Philip J Poole, Val Zwiller, Klaus D Jöns, On-chip single photon filtering and multiplexing in hybrid quantum photonic circuits, \textit{Nature communications}, \textbf{8}(1), 1--8 (2017).
    
    \bibitem{zadeh2016deterministic}
    Iman Esmaeil Zadeh, Ali W Elshaari, Klaus D Jons, Andreas Fognini, Dan Dalacu, Philip J Poole, Michael E Reimer, Val Zwiller, Deterministic integration of single photon sources in silicon based photonic circuits, \textit{Nano Letters}, \textbf{16}(4), 2289--2294 (2016).
    
    \bibitem{elshaari2018strain}
    Ali W Elshaari, Efe Buyukozer, Iman Esmaeil Zadeh, Thomas Lettner, Peng Zhao, Eva Scholl, Samuel Gyger, Michael E Reimer, Dan Dalacu, Philip J Poole, Strain-tunable quantum integrated photonics, \textit{Nano letters}, \textbf{18}(12), 7969--7976 (2018).
    
    \bibitem{gourgues2019controlled}
    Ronan Gourgues, Iman Esmaeil Zadeh, Ali W Elshaari, Gabriele Bulgarini, Johannes WN Los, Julien Zichi, Dan Dalacu, Philip J Poole, Sander N Dorenbos, Val Zwiller, Controlled integration of selected detectors and emitters in photonic integrated circuits, \textit{Optics express}, \textbf{27}(3), 3710--3716 (2019).
    
    \bibitem{pan2000experimental}
    Jian-Wei Pan, Dik Bouwmeester, Matthew Daniell, Harald Weinfurter, Anton Zeilinger, Experimental test of quantum nonlocality in three-photon Greenberger--Horne--Zeilinger entanglement, \textit{Nature}, \textbf{403}(6769), 515--519 (2000).
    
    \bibitem{lombardi2002teleportation}
    Egilberto Lombardi, Fabio Sciarrino, Sandu Popescu, Francesco De Martini, Teleportation of a vacuum--one-photon qubit, \textit{Physical review letters}, \textbf{88}(7), 070402 (2002).
    
    \bibitem{qiao2021multistage}
    Lu-Feng Qiao, Zhi-Qiang Jiao, Xiao-Yun Xu, Jun Gao, Zhe-Yong Zhang, Ruo-Jing Ren, Wen-Hao Zhou, Xiao-Wei Wang, Xian-Min Jin, Multistage quantum swapping of vacuum-one-photon entanglement, \textit{Physical Review A}, \textbf{104}(2), 022415 (2021).
    
    \bibitem{das2022optimal}
    Tamoghna Das, Marcin Karczewski, Antonio Mandarino, Marcin Markiewicz, Marek {.Z}ukowski, Optimal Interferometry for Bell Nonclassicality Induced by a Vacuum--One-Photon Qubit, \textit{Physical Review Applied}, \textbf{18}(3), 034074 (2022).
    
    \bibitem{chitambar2014}
    Eric Chitambar, Debbie Leung, Laura Man{\v{c}}inska, Maris Ozols, Andreas Winter, Everything you always wanted to know about LOCC (but were afraid to ask), \textit{Communications in Mathematical Physics}, \textbf{328}(1), 303--326 (2014).
    
    \bibitem{nielsen1999}
    Michael A Nielsen, Conditions for a class of entanglement transformations, \textit{Physical Review Letters}, \textbf{83}(2), 436 (1999).
    
    \bibitem{vicente2013}
    Julio I de Vicente, Cornelia Spee, Barbara Kraus, Maximally entangled set of multipartite quantum states, \textit{Physical review letters}, \textbf{111}(11), 110502 (2013).
    
    \bibitem{mermin1990}
    N David Mermin, Extreme quantum entanglement in a superposition of macroscopically distinct states, \textit{Physical Review Letters}, \textbf{65}(15), 1838 (1990).
    
    \bibitem{greenberger1990}
    Daniel M Greenberger, Michael A Horne, Abner Shimony, Anton Zeilinger, Bell’s theorem without inequalities, \textit{American Journal of Physics}, \textbf{58}(12), 1131--1143 (1990).
    
    \bibitem{wei2003}
    Tzu-Chieh Wei, Paul M Goldbart, Geometric measure of entanglement and applications to bipartite and multipartite quantum states, \textit{Physical Review A}, \textbf{68}(4), 042307 (2003).
    
    \bibitem{steinberg2022}
    Jonathan Steinberg, Otfried Guhne, Maximizing the geometric measure of entanglement, \textit{arXiv preprint arXiv:2210.13475}, [Online; accessed 24-Oct -2022].
    
    \bibitem{wang2004image}
    Zhou Wang, Alan C Bovik, Hamid R Sheikh, Eero P Simoncelli, Image quality assessment: from error visibility to structural similarity, \textit{IEEE transactions on image processing}, \textbf{13}(4), 600--612 (2004).
    
    \bibitem{chang2023nanowire}
    Jin Chang, Jun Gao, Iman Esmaeil Zadeh, Ali W Elshaari, Val Zwiller, Nanowire-based integrated photonics for quantum information and quantum sensing, \textit{Nanophotonics}, \textbf{12}(3), 339--358 (2023).
    
    \bibitem{esmaeil2021superconducting}
    Iman Esmaeil Zadeh, J Chang, Johannes WN Los, Samuel Gyger, Ali W Elshaari, Sander N Dorenbos, Val Zwiller, Superconducting nanowire single-photon detectors: A perspective on evolution, state-of-the-art, future developments, and applications, \textit{Applied Physics Letters}, \textbf{118}(19), 190502 (2021).
    
    \bibitem{esmaeil2020efficient}
    Iman Esmaeil Zadeh, Johannes WN Los, Ronan BM Gourgues, Jin Chang, Ali W Elshaari, Julien Romain Zichi, Yuri J van Staaden, Jeroen PE Swens, Nima Kalhor, Antonio Guardiani, others, Efficient single-photon detection with 7.7 ps time resolution for photon-correlation measurements, \textit{Acs Photonics}, \textbf{7}(7), 1780--1787 (2020).
    
    \bibitem{moody20222022}
    Galan Moody, Volker J Sorger, Daniel J Blumenthal, Paul W Juodawlkis, William Loh, Cheryl Sorace-Agaskar, Alex E Jones, Krishna C Balram, Jonathan CF Matthews, Anthony Laing, others, 2022 Roadmap on integrated quantum photonics, \textit{Journal of Physics: Photonics}, \textbf{4}(1), 012501 (2022).
    

\end{thebibliography}
\end{document}